# Penambahan Emosi Menggunakan Metode Manipulasi Prosodi Untuk Sistem *Text To Speech* Bahasa Indonesia

Salita Ulitia. P[1], Ary S. Prihatmanto[2]

School of Electrical Engineering and Informatics, Institut Teknologi Bandung, Bandung, Indonesia

[1]salitaulitia@gmail.com, [2]asetijadi@lskk.ee.itb.ac.id

*Abstrak*—***Text To Speech*** **(TTS) merupakan suatu sistem yang dapat mengonversi teks dalam format suatu bahasa menjadi ucapan sesuai dengan pembacaan teks dalam bahasa yang digunakan. Fokus penelitian yaitu suatu konsep pengucapan natural, dengan usaha "memanusiakan" pelafalan sintesa suara sistem *Text To Speech* yang dihasilkan. Kebutuhan utama yang digunakan untuk sistem *Text To Speech* dalam penelitian ini adalah eSpeak, database MBROLA id1, *database Human Speech Corpus* diambil dari suatu *website* yang merangkum kata-kata dengan frekuensi terbanyak *(Most Common Words)* yang digunakan pada suatu negara, dan terdapat 3 jenis emosi/intonasi dasar yang dirancang, yaitu emosi senang (*happy*), marah (*angry*), dan sedih (*sad*). Pendekatan metode yang digunakan untuk mengembangkan filter emosional adalah dengan memanipulasi fitur yang relevan dari prosodi (terutama nilai *pitch* dan durasi) menggunakan suatu *predetermined rate factor* yang telah ditetapkan. Hasil pengujian tes persepsi *Human Speech Corpus* adalah untuk emosi senang sebesar 95 %, emosi marah sebesar 96.25 % dan emosi sedih sebesar 98.75 %. Untuk aspek uji *intelligibility* ketepatan suara yang didengar dengan kalimat asli adalah sebesar 93.3 %, dan untuk rate kejelasan untuk masing-masing kalimat adalah 62.8 %. Untuk aspek uji *naturalness* ketepatan pemilihan emosi adalah sebesar 75.6 % dengan masing-masing emosi senang sebesar 90 %, emosi marah sebesar 73.3 % dan emosi sedih sebesar 60 %.**

***Kata kunci*; *Text To Speech*, eSpeak, MBROLA*, Human Speech Corpus*, emosi, intonasi, manipulasi prosodi.**

I. Pendahuluan

Perkembangan interaksi antara manusia dan komputer telah mengalami kemajuan dari waktu ke waktu. *Text To Speech* (TTS) merupakan suatu sistem yang dapat mengonversi teks dalam format suatu bahasa menjadi ucapan sesuai dengan pembacaan teks dalam bahasa yang digunakan. Suara yang merupakan keluaran dari TTS akan mempermudah seseorang mempelajari pengucapan suatu kata dalam bahasa tertentu, karena setiap bahasa memiliki keunikan dan aturan pengucapan yang berbeda untuk setiap bahasa.

Pada masa ini, setelah sistem *Text To Speech* menjadi hal yang lazim diaplikasikan dalam kehidupan sehari-hari, terdapat tantangan agar sistem *Text To Speech* tersebut menjadi suatu sistem yang memiliki kemampuan untuk memberikan keakuratan prosodi atau lafal pengucapan yang dapat dimengerti oleh pendengar agar mengerti ucapan yang disintesa.

Fokus penelitian ini yaitu suatu konsep pengucapan natural, dengan usaha "memanusiakan" pelafalan sintesa suara sistem *Text To Speech* yang dihasilkan. Manusia memiliki emosi/intonasi yang dapat mempengaruhi suara yang dihasilkan. Hal inilah yang mendasari emosi/intonasi menjadi fokus utama penelitian. Penyampaian emosi yang tepat merupakan hal yang sangat penting dalam sistem ucapan, seperti aplikasi penerjemah, dimana nada sumber pengucap, dan kalimat yang diucapkan harus sesuai agar tujuan pengucapan dapat tersampaikan dengan baik.

II. Tinjauan Pustaka

*A. Tinjauan Text To Speech (TTS)*

Sistem teks ke ucapan (*Text To Speech*), diperlukan untuk mengubah teks yang dihasilkan oleh komputer menjadi ucapan. TTS sering disebut dengan pensintesa ucapan atau *Speech Synthesizer* [8].

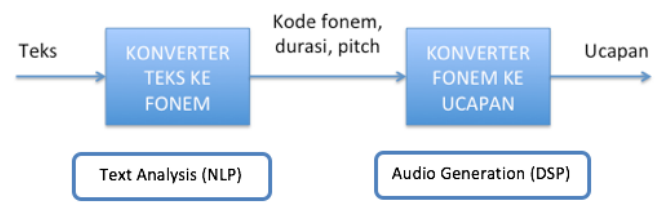

Gambar 1. Sistem *Text To Speech*.

Bagian konverter teks ke fonem berfungsi untuk mengolah kalimat masukan dalam suatu bahasa tertentu yang berbentuk teks menjadi urutan kode-kode bunyi yang direpresentasikan dengan kode fonem, durasi serta *pitch*. Bagian konverter fonem ke ucapan menerima masukan kode-kode fonem serta *pitch* dan durasi yang telah dihasilkan oleh bagian sebelumnya. Berdasarkan kode-kode tersebut, bagian ini akan menghasilkan bunyi atau sinyal ucapan yang sesuai dengan kalimat masukan.

*B. Tinjauan eSpeak*

eSpeak [2] adalah suatu perangkat lunak *speech synthesizer* yang terkonfigurasi dengan berbagai macam bahasa untuk sistem operasi Linux dan Windows.

eSpeak tersedia pada suatu program *command line* (Linux dan Windows) untuk mengucapkan teks dari suatu file atau *stdin* (*input keyboard*), suatu perpustakaan bersama yang dapat digunakan oleh program lain (pada Windows berupa DLL), dan suatu versi SAPI5 untuk Windows, sehingga dapat digunakan dengan pembaca layar atau program lain yang mendukung *interface* Windows SAPI5.

*C. Tinjauan MBROLA*

MBROLA *project* [3] adalah suatu kumpulan dari *diphone voice* untuk sintesa speech. MBROLA dikembangkan di TCTS *Laboratory (Circuit Theory and Signal Processing Laboratory)* of The Faculté Polytechnique de Mons (Belgium) dengan tujuan untuk menghasilkan suatu set *speech synthesizer* untuk sebanyak mungkin bahasa dan untuk pengembangan selanjutnya dengan tujuan non-komersil.

Pusat pengembangan MBROLA adalah suatu *speech synthesizer* berdasarkan gabungan dari *diphone*. Dibutuhkan daftar fonem sebagai input, bersamaan dengan informasi prosodi (durasi fonem dan deskripsi *pitch*), dan menghasilkan *sample* sebesar 16 bits.

eSpeak dapat digunakan sebagai aplikasi yang dapat menggunakan MBROLA sebagai diphone voicenya. eSpeak menyediakan fitur penerjemah dan intonasi *spelling-to-phonem* MBROLA tersebut sehingga dapat menghasilkan suara sesuai voice yang dipilih.

Untuk menggunakan MBROLA voice, eSpeak membutuhkan informasi untuk menerjemahkan dari fonemnya sendiri, tetapi ekuivalen dengan fonem MBROLA. eSpeak yang menggunakan MBROLA voice diberikan nama : mb-xxx, dimana xxx merupakan nama dari MBROLA voice. File voice ini terdapat di direktori espeak-data/voices/mbrola. (eg. mb-id1 untuk MBROLA "id1" Indonesia).

*D. Tinjauan Korpus*

Menurut Kamus Besar Bahasa Indonesia (KBBI), korpus data adalah sekumpulan data yang digunakan sebagai sumber bahan penelitian. Istilah korpus saat ini paling banyak digunakan untuk merujuk kepada sekumpulan data linguistik yang dikumpulkan untuk tujuan analitik tertentu, untuk selanjutnya disimpan, dikelola, dan dianalisis dalam bentuk digital.

Untuk sistem *Text To Speech* pada penelitian tesis ini, database yang digunakan berdasarkan pada MBROLA *voices* yang dipakai, yaitu bahasa Indonesia (Id). Sedangkan untuk *Human Speech Corpus* diambil dari suatu *website* http://www.ezglot.com/most-frequently-used-words.php [11] dan didapatkan kurang lebih 2000 kata yang paling banyak digunakan pada suatu negara, dan dapat dimanfaatkan untuk kebutuhan *library database* yang akan dibuat. Database ini berisi penggalan suku kata-kata dalam SAMPA (*Speech Assessment Methods Phonetic Alphabet*) [9].

*E. Tinjauan Emosi/Intonasi*

Emosi dapat didefinisikan sebagai suatu perubahan dalam keadaan kesiapan untuk mempertahankan atau memodifikasi hubungan dalam lingkungan [12]. Emosi manusia tidak hanya mempengaruhi gerakan tubuh dan aktivitas otak akan tetapi juga mempengaruhi cara manusia berkomunikasi. Komunikasi manusia lebih dari sekedar kata-kata. Dalam komunikasi manusia, kata-kata hanya mencapai 7% dalam komunikasi manusia dan 38% diwakili oleh kualitas suara, ekspresi vokal dan prosodi. Sisanya 55% merupakan kontribusi faktor-faktor lain terutama wajah ekspresi dan tubuh isyarat [18].

Dalam mengembangkan sistem TTS menggunakan emosi/intonasi, fokusnya adalah pada aspek prosodi karena manusia menyampaikan emosi mereka dalam ucapan dengan memanipulasi prosodi termasuk variasi dalam ucapan, tempo dan tingkat *stress*.

Tabel 1. Pengaruh emosi pada suara yang dihasilkan oleh manusia.

| | Fear | Anger | Sorrow | Joy | Disgust | Surprise |
|---|---|---|---|---|---|---|
| Speech rate | Much faster | Slightly faster | Slightly slower | Faster and slower | Very much slower | Much faster |
| Pitch average | Very much higher | Very much higher | Slightly lower | Much higher | Very much lower | Much higher |
| Pitch range | Much wider | Much wider | Slightly narrower | Much wider | Slightly wider | |
| Intensity | Normal | Higher | Lower | Higher | lower | Higher |
| Voice quality | Irregular voicing | Breathy chest tone | resonant | Breathy blaring | Grumbled chest tone | |
| Pitch changes | Normal | Abrupt on stressed syllable | Downward inflections | Smooth upward inflections | Wide downward terminal inflections | Rising contour |
| Articulation | Precise | Tense | | Normal | normal | |

## III. DESAIN DAN PERANCANGAN SISTEM

*A. Desain Sistem Text To Speech*

Sistem *Text To Speech* yang akan dibuat sama seperti IndoTTS yang dikembangkan oleh Arry Akhmad Arman [6], akan tetapi *Text To Speech* yang dibuat dalam penelitian tesis ini menggunakan *engine* eSpeak dan database *synthesizer* MBROLA, yang menggunakan metode *diphone concatenation* untuk mensintesis suara

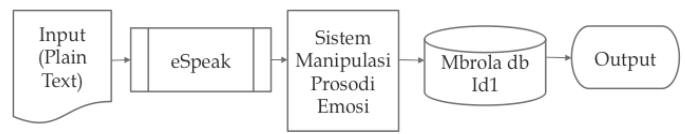

Gambar 2. Diagram blok sistem *Text To Speech* yang dirancang.

Terdapat 3 proses utama pada Text To Speech, yaitu normalizer, phonetizer, dan prosodi. Normalizer adalah tahap pemisahan kata-kata menjadi fonem tunggal, phonetizer adalah tahap mengonversi fonem tunggal menjadi karakter yang diakui oleh MBROLA, dan penambahan prosodi, yaitu nilai durasi dan nilai *pitch* pada fonem tunggal tersebut.

Gambar diagram fungsional sistem Text To Speech dapat dilihat pada gambar 3 berikut.

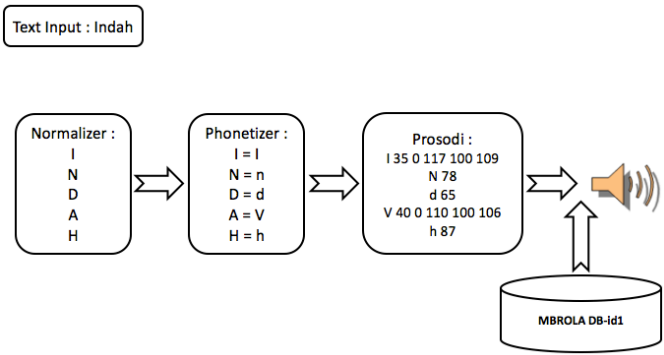

Gambar 3. Diagram fungsional sistem *Text To Speech*.

### B. Metode Manipulasi Prosodi

Pendekatan metode yang digunakan untuk mengembangkan filter emosional adalah dengan memanipulasi fitur yang relevan dari prosodi (terutama *pitch* dan durasi) menggunakan suatu *predetermined rate factor* yang telah ditetapkan dengan menganalisis perbedaan antara output standar *Text To Speech* dan prosodi perekaman suara dengan emosi/intonasi tertentu [1].

Pendekatan ini diambil karena tidak memerlukan suatu *large emotional* database untuk meminimalisasi sumber daya yang dapat digunakan. Yang dibutuhkan adalah suatu algoritma yang dapat memanipulasi *pitch* dan durasi output.

Setiap fonem diberikan karakteristik prosodi, kualitas suara dan artikulasi. Metode ini berlaku untuk penambahan atau pengurangan untuk *pitch*, durasi, kualitas suara dan vokal presisi. Peningkatan dan pengurangan dapat diterapkan untuk fonem tertentu, suku kata atau seluruh ucapan. Tahap metode manipulasi prosodi dapat dilihat pada Gambar 4 berikut.

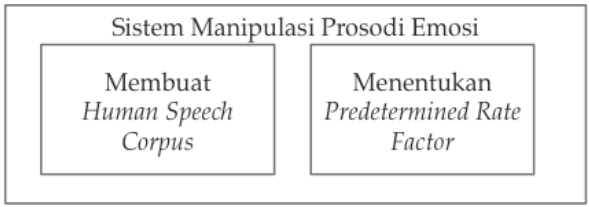

Gambar 4. Diagram fungsional sistem *Text To Speech*.

Sistem terdiri atas pembuatan *human speech corpus* dan menentukan *predetermined rate factor*. *Human speech corpus* dibuat untuk membantu memahami ucapan dan kaitannya dengan emosi/intonasi manusia. Tahapan dari pembuatan *human speech corpus* ini diantaranya adalah membuat kalimat berdasarkan pola suku kata sesuai emosi yang akan diberikan, sesi perekaman kalimat setiap emosi, dan tes persepsi.

Tabel 2. Contoh kalimat yang digunakan untuk perekaman.

| Emosi | Struktur Suku Kata | Kata Pertama | Kata Kedua | Kata Ketiga |
|---|---|---|---|---|
| Senang | 223 | A ku | Su ka | Se ka li |
| Senang | 322 | Se ni or | A mat | Can tik |
| Marah | 222 | Ka mu | Di am | Sa ja |
| Marah | 233 | Per gi | Ka lian | Ber du a |
| Sedih | 232 | Hi lang | Ha di ah | I tu |
| Sedih | 322 | Lu pa kan | Sa ja | A ku |

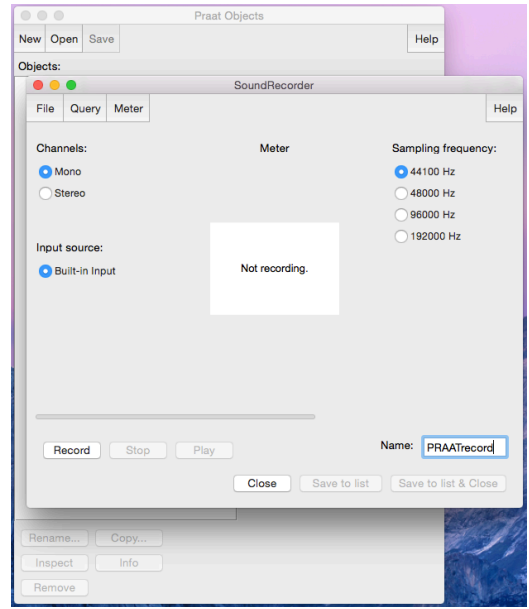

Gambar 5. Diagram fungsional sistem *Text To Speech*.

Tes persepsi ini penting sebelum melanjutkan ke tahapan analisis untuk mendapatkan *predetermined rate factor*. Formulir tes persepsi adalah pada Gambar 6 berikut.

Gambar 6. Formulir tes persepsi rekaman *Human Speech Corpus*.

Sesuai (Lisombe et al., 2003) [11], nilai persentase minimum rekognisi adalah 60%. Apabila persentasenya belum mencapai yang telah ditetapkan, akan dilakukan penggantian kalimat dengan tahap seperti yang telah dilakukan sebelumnya.

Hasil persepsi *Human Speech Corpus* yang telah dirancang dapat dilihat pada Tabel 3.

Tabel 3. Hasil tes persepsi *Human Speech Corpus*.

| Kalimat | Pilihan Emosi Listeners | | |
|---|---|---|---|
| | SENANG | MARAH | SEDIH |
| SENANG | **95 %** | 5 % | 0 % |
| MARAH | 2.5 % | **96.25 %** | 1.25 % |
| SEDIH | 0 % | 1.25 % | **98.75 %** |

Tahap menentukan *predetermined rate factor* adalah segmentasi per fonem untuk durasi dan *pitch* dengan *software* Praat [10] untuk kemudian dibandingkan dengan *output* standar TTS, lalu dilakukan perhitungan nilai durasi dan *pitch* per suku kata.

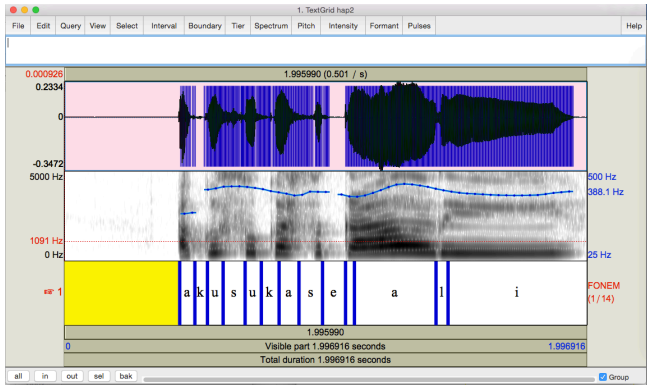

Gambar 7. Proses segmentasi kalimat dengan *software* Praat.

Rumus perhitungan *predetermined rate factor* adalah sebagai berikut [1].

$$Duration\ diff = \frac{selisih\ durasi\ TTS\ standar\ dan\ Praat}{durasi\ TTS\ standar} x100$$

$$Starting\ Pitch\ diff = \frac{selisih\ starting\ pitch\ TTS\ standar\ dan\ Praat}{starting\ pitch\ TTS\ standar} x100$$

$$Ending\ Pitch\ diff = \frac{selisih\ ending\ pitch\ TTS\ standar\ dan\ Praat}{ending\ pitch\ TTS\ standar} x100$$

Boolean untuk nilai (+) dan (−) pada duration dan pitch diff adalah:
apabila nilai duration atau pitch > TTS standar maka nilai (+)
apabila nilai duration atau pitch < TTS standar maka nilai (−)

Tabel 4. Analisis durasi untuk emosi senang.

| Kata | Suku Kata | Konsonan Pertama | Konsonan Kedua | Vokal |
|---|---|---|---|---|
| Kata Pertama | Awal | (-)38% | (-)32% | (+)65% |
| | Tengah | (+)4% | (-)36% | (+)61% |
| | Akhir | (-)11% | (+)30% | (+)89% |
| Kata Terakhir | Awal | (-)16% | (+)53% | (+)174% |
| | Tengah | (-)5% | (+)21% | (+)252% |
| | Akhir | (+)14% | (+)60% | (+)256% |
| Kata Lain | | (+)21% | | |

Tabel 5. Analisis *pitch* untuk emosi senang.

| Kata | Suku Kata | Pitch Awal | Pitch Akhir |
|---|---|---|---|
| Kata Pertama | Awal | (+)132% | (+)152% |
| | Tengah | (+)133% | (+)162% |
| | Akhir | (+)173% | (+)201% |
| Kata Terakhir | Awal | (+)205% | (+)282% |
| | Tengah | (+)243% | (+)363% |
| | Akhir | (+)288% | (+)333% |
| Kata Lain | | (+)226% | (+)242% |

Tabel 6. Analisis durasi untuk emosi marah.

| Kata | Suku Kata | Konsonan Pertama | Konsonan Kedua | Vokal |
|---|---|---|---|---|
| Kata Pertama | Awal | (-)36% | (+)1% | (+)12% |
| | Tengah | (+)5% | (+)3% | (+)72% |
| | Akhir | (-)15% | (+)5% | (+)136% |
| Kata Terakhir | Awal | (-)18% | (-)10% | (+)77% |
| | Tengah | (-)41% | (+)5% | (+)26% |
| | Akhir | (+)9% | (-)16% | (+)58% |
| Kata Lain | | (+)5% | | |

Tabel 7. Analisis *pitch* untuk emosi marah.

| Kata | Suku Kata | Pitch Awal | Pitch Akhir |
|---|---|---|---|
| Kata Pertama | Awal | (+)172% | (+)183% |
| | Tengah | (+)195% | (+)260% |
| | Akhir | (+)244% | (+)237% |
| Kata Terakhir | Awal | (+)192% | (+)232% |
| | Tengah | (+)154% | (+)207% |
| | Akhir | (+)209% | (+)202% |
| Kata Lain | | (+)192% | (+)205% |

Tabel 8. Analisis durasi untuk emosi sedih.

| Kata | Suku Kata | Konsonan Pertama | Konsonan Kedua | Vokal |
|---|---|---|---|---|
| Kata Pertama | Awal | (-)25% | (-)40% | (+)113% |
| | Tengah | (-)11% | (+)40% | (+)135% |
| | Akhir | (+)8% | (+)26% | (+)40% |
| Kata Terakhir | Awal | (+)10% | (-)7% | (+)117% |
| | Tengah | (+)8% | (+)33% | (+)169% |
| | Akhir | (+)10% | (+)10% | (+)229% |
| Kata Lain | | (+)33% | | |

Tabel 9. Analisis *pitch* untuk emosi sedih.

| Kata | Suku Kata | Pitch Awal | Pitch Akhir |
|---|---|---|---|
| Kata Pertama | Awal | (+)123% | (+)122% |
| | Tengah | (+)96% | (+)108% |
| | Akhir | (+)107% | (+)112% |
| Kata Terakhir | Awal | (+)88% | (+)133% |
| | Tengah | (+)89% | (+)137% |
| | Akhir | (+)116% | (+)136% |
| Kata Lain | | (+)113% | (+)116% |

C. Implementasi Metode Manipulasi Prosodi

Implementasi dari metode manipulasi prosodi untuk sistem *Text To Speech* yang dirancang dapat dilihat pada Gambar 8.

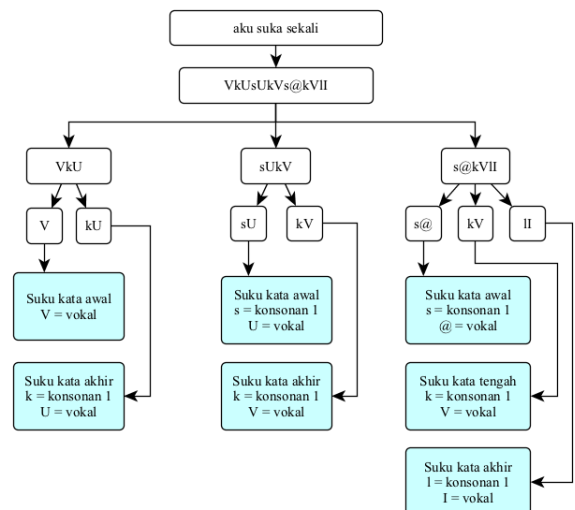

Gambar 8. Bagan implementasi manipulasi prosodi.

Algoritma dari bagan implementasi manipulasi prosodi diatas adalah :

1) Menginput kalimat, espeak men-*generate* standar *phonem* di konsole
2) Sistem membaca seluruh SAMPA pada *phonem* beserta durasi dan pitch kemudian disimpan sementara dalam *database*
3) Sistem mengecek setiap kata yang diinputkan kemudian dibandingkan dengan *database* kata dalam bentuk *phonem* SAMPA untuk mendapatkan indeks kata
4) Sistem mengecek kata pada *database* SAMPA untuk mengetahui letak indeks
5) Kemudian melihat di *database* suku kata dan suku sampa untuk manipulasi nilai *pitch* dan durasi dari *predeterminated rate factor*
6) Manipulasi nilai sesuai aturan suku kata
7) Membuat pho baru hasil manipulasi
8) SAMPA *phoneme* pada *database* dihapus.

## IV. Pengujian dan Analisis

### A. Desain Sistem yang Telah Dibuat

Tampilan program sistem Text To Speech penambahan emosi/intonasi dengan metode manipulasi prosodi tersebut dapat dilihat pada Gambar 9 berikut.

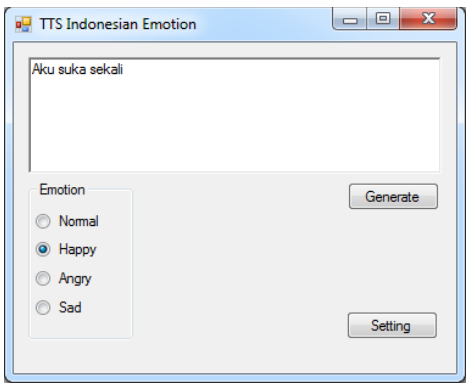

Gambar 9. Tampilan program sistem *Text To Speech*.

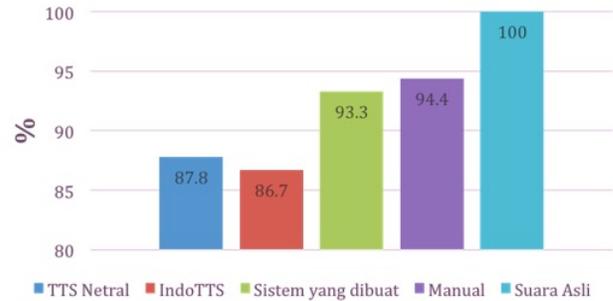

Gambar 10. Tampilan *predetermined rate factor* pada sistem *Text To Speech*.

Pengujian dilakukan dengan mempertimbangkan 2 jenis aspek uji i*ntelligibility* (kejelasan) *dan naturalness* (kealamian atau kenaturalan).

Untuk aspek *intelligibility*, *listeners* akan mendengarkan kalimat yang sebelumnya telah direkam untuk setiap sistem dalam format *.wav* secara acak, kemudian diminta menulis kalimat yang didengar dan memberikan *rate* kejelasan kalimat yang didengar, mulai dari amat jelas hingga amat buruk.

Untuk aspek *naturalness*, *listeners* juga akan mendengarkan kalimat yang sebelumnya telah direkam untuk setiap sistem dalam format *.wav* secara acak, kemudian diminta memilih emosi sesuai yang dikenalinya.

### B. Hasil Pengujian dan Analisa

Grafik presentase untuk masing-masing uji *intelligibility* dan *naturalness* diperlihatkan pada gambar berikut.

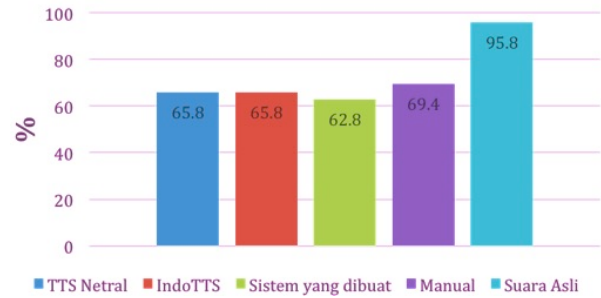

Gambar 11. Grafik persentase rata-rata hasil uji *intelligibility* menuliskan setiap kalimat yang didengar.

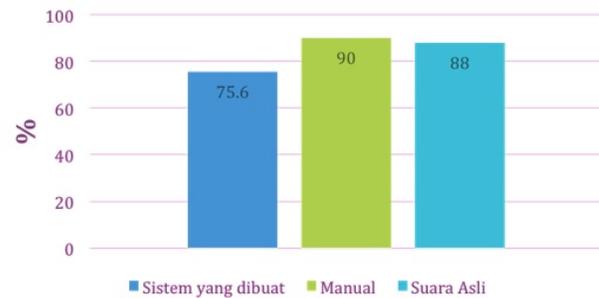

Gambar 12. Grafik persentase rata-rata hasil uji *intelligibility* memberikan *rate* kejelasan setiap kalimat yang didengar.

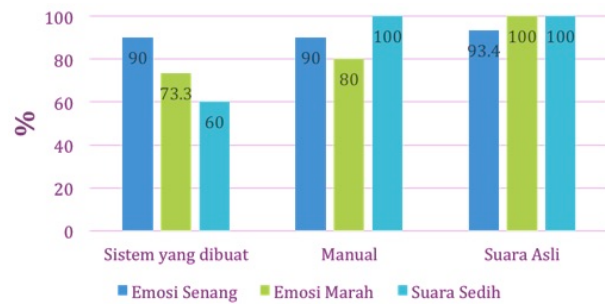

Gambar 13. Grafik persentase rata-rata hasil uji *naturalness* ketepatan memilih emosi setiap kalimat.

Gambar 14. Grafik persentase rata-rata hasil uji *naturalness* rekognisi masing-masing emosi.

Hasil pengujian tes persepsi untuk aspek uji *intelligibility* memperlihatkan bahwa kecenderungan *listeners* tidak mudah untuk mendengar suara hasil sintesa *(Text To Speech)* dibandingkan suara asli.

Untuk hasil pengujian tes persepsi untuk aspek uji *naturalness* memperlihatkan *listeners* mampu memilih emosi sesuai dengan emosi pada sistem.

## V. KESIMPULAN DAN SARAN

### A. Kesimpulan

1. Penerapan model manipulasi prosodi pada sistem *Text To Speech* yang dirancang dapat dilakukan sesuai fungsinya.
2. Emosi/intonasi yang berhasil diterapkan dalam sistem *Text To Speech* yang dirancang adalah berupa emosi dasar manusia yaitu emosi senang (*happy*), marah (*angry*), dan sedih (*sad*).
3. Hasil pengujian tes persepsi untuk *Human Speech Corpus* adalah untuk emosi senang sebesar 95 %, emosi marah sebesar 96.25 % dan emosi sedih sebesar 98.75%.
4. Pengujian tes persepsi untuk sistem *Text To Speech* dilakukan dengan aspek uji *intelligibility* dan aspek uji *naturalness*. Untuk aspek uji *intelligibility* ketepatan suara yang didengar dengan kalimat asli adalah sebesar 93.3 %, dan untuk *rate* kejelasan untuk masing-masing kalimat adalah 62.8 %. Untuk aspek uji *naturalness* ketepatan pemilihan emosi adalah sebesar 75.6 % dengan rekognisi masing-masing emosi senang sebesar 90 %, emosi marah sebesar 73.3 % dan emosi sedih sebesar 60 %.

### B. Saran

1. Penambahan *database* kata-kata yang digunakan untuk pengucapan, terutama untuk kata berimbuhan. Dapat menambah dengan seluruh kata-kata dalam kamus seperti KBBI (Kamus Besar Bahasa Indonesia) dan/atau *database* korpus lain.
2. Memperbanyak kalimat uji untuk pengujian sampel *Human Speech Corpus* agar nilai dari analisis yang dilakukan mampu mendekati atau mewakili emosi sebenarnya.
3. Penambahan jenis emosi/intonasi yang lebih spesifik, seperti takut (*fear*), bosan (*boredom*), putus asa (*despair*), frustration (*kekecewaan*), dan lain sebagainya.
4. Penambahan aksen atau logat untuk lafal pengucapan emosi/intonasi. Seperti marah dalam logat Sunda, sedih dalam logat Jawa, senang dalam logat Bugis, dan lain sebagainya.
5. Apabila memungkinkan, dapat menggunakan bahasa lain seperti Inggris, Perancis, dan lain sebagainya dengan terlebih dahulu mempelajari pola suku kata bahasa tersebut.